\documentclass[aps,prd,twocolumn,superscriptaddress,amsmath,amssymb]{revtex4}
\pdfoutput=1
\usepackage{graphicx}
\usepackage{color}
\usepackage{ulem}
\usepackage{hyperref}
\usepackage{float}
\usepackage{cancel}
\begin{document}
	\title{ Probing CPT breaking induced by quantum decoherence at DUNE}
	\author{J.C. Carrasco}
	\author{F.N. D\'iaz}
	\author{A.M. Gago}
	\affiliation{Secci\'on F\'isica, Departamento de Ciencias, Pontificia Universidad Cat\'olica del Per\'u, Apartado 1761, Lima, Per\'u}	
	\begin{abstract}
We show that, the decoherence phenomena applied to the neutrino system could lead us to have an observable breaking of the fundamental CPT symmetry. 
We require a specific textures of non-diagonal decoherence matrices, with non-zero $\delta_{CP}$, for having such observations. Using the information from the CPT conjugate channels: $\nu_{\mu} \rightarrow \nu_{\mu}$ and $\bar{\nu}_{\mu} \rightarrow \bar{\nu}_{\mu}$ 
and its corresponding backgrounds, we have estimated the sensitivity of DUNE experiment for testing CPT under the previous conditions. Four scenarios for energy dependent decoherence parameters $\Gamma_{E_\nu}=\Gamma \times (E_\nu/\mathrm{GeV})^n$, $n=-1,0,1,$ and $2$ are taken into account, for most of them, DUNE is able to achieve a 5$\sigma$ discovery potential having $\Gamma$ in $\mathcal{O} (10^{-23}$ GeV) for $\delta_{CP}=3\pi/2$. Meanwhile, for $\delta_{CP}=\pi/2$ we reach 3$\sigma$ for $\Gamma$ in $\mathcal{O} (10^{-24}$ GeV).      

	\end{abstract}
	\maketitle
	
	\section{INTRODUCTION}

The neutrino oscillation is
provoked by the existence of non-zero neutrino masses allied with the mismatch
between its corresponding eigenstates and the neutrino flavor eigenstates. This phenomenon is supported by an overwhelming experimental evidence which spans more than two decades ago \cite{Fukuda01,Ahmad02,Fukuda98,Kajita16,Araki05,An12,Adamson14,Ahn12,Abe12,McDonald05}. Notwithstanding, the neutrino oscillation is well-established, the coexistance of new physics as sub-leading effect of it has not been yet rule out. In some occasions, this new physics bring about the option of  breaking fundamental laws of nature, for instance, the violation of the equivalence principle \cite{Gasperini88, Halprin91, Gago:1999hi} or the violation of lorentz invariance \cite{Colladay:1996iz, Coleman:1998ti, Coleman:1997xq}. The latter is described within the Lagrangian of the Standard Model Extension (SME) \cite{Colladay:1998fq} where you can find terms that explicitly violate the combined action of the conjugation (C), the parity inversion (P) and the time inversion(T), symmetries, known in short as the CPT symmetry. This combined symmetry holds for a local, lorentz invariant and unitary quantum field theory. There has been a lot of work on testing CPT violation (CPTV) on the side of the SME \cite{Adamson:2008aa,Adamson:2010rn, 
AguilarArevalo:2011yi,Li:2014rya}. We must remark that, there is CPTV in the neutrino oscillation in matter, originated by the unequal number of particles and antiparticles in ordinary matter \cite{Jacobson:2003wc}.

On the other hand, there are a set of theoretical hypotheses such as string and branes~\cite{Ellis,Benattistrings}, and quantum gravity~\cite{Hawking1} which effects can be encoded behind an omnipresent environment that could be weakly interacting with neutrinos \cite{Benatti00,Benatti01}. This type of interactions is written according to the open system formalism and have as a typical trait (in its simplistic version) the appearance of decoherence (damping) factors $\exp^{-\Gamma t}$ within the neutrino oscillation probabilities \cite{Lisi00,Gago01a,Gago02a, Barenboim:2004wu, Fogli07, Farzan08,Oliveira13,Carpio:2017nui}. Here, arises other type of CPTV  rooted in the impossibility of defining a CPT operator by virtue of the evolution from pure states to mixed states, caused by decoherence~\cite{Wald80,Mavromatos:2009ww}. In the open system approach the effects of the environment are enclosed, in a model independent way, in the so-called dissipative/decoherence matrix (after tracing out the environment degrees of freedom). Thereby, it is uncertain to claim that this type of CPTV is genuine, i.e. there is a fundamental arrow of time, or it is only an apparent CPTV, in view of our lack of knowledge of the complete system. One way or another, an eventual observation of CPTV will be shaking our current understanding of fundamental physics.

In this paper, we will focus on study different non-diagonals textures of the dissipative matrix paying special attention to those which can produce an observable non-zero CPTV. An equal response of the environment for neutrinos and antineutrinos will be one of our working hypothesis. In constrast with the hypothesis use, for instance, in~\cite{Barenboim:2004wu}. We will also consider the possibility that the parameters of the dissipative matrix can be energy dependent~\cite{Liu97, EllisMavro96}. 
We will use the DUNE experiment ~\cite{cdr,ancillaryfiles} as the scenario for assesing how significant would be a CPTV signature caused by quantum decoherence.

	\section{Theorical approach}
	\subsection{Neutrino as open quantum system}
	\label{DensityMatrixFormalism}
Our aim is to treat the neutrino as a subsystem interacting, weakly, with a large (unknown) environment.  In situations of this kind, the linear evolution of the reduced density matrix of the subsystem is represented by means of the Lindblad Master equation~\cite{Benatti00,Benatti01}:	
	\begin{equation}
	\label{Lindblad}
	\frac{\partial\rho(t)}{\partial t}=-i[H,\rho(t)]+\mathcal{D}[\rho(t)],
	\end{equation}
	where $\rho(t)$ is the neutrino density matrix, $H$
	is the hamiltonian of the neutrino subsystem and $\mathcal{D}[\rho(t)]$ is the dissipative term where the decoherence phenomena is encoded. This dissipative factor is written as follows:
	\begin{equation}
	\label{DissipativeTerm}
	\mathcal{D}[\rho(t)]=\frac{1}{2}\sum_{j} \left([A_j,\rho(t) A_j^{\dagger}]+[A_j\rho(t), A_j^{\dagger}] \right).
	\end{equation}
Considering a three-level system we can expand the operators in Eq.~(\ref{Lindblad}) in the basis of the Gell-Mann matrices from $SU(3)$ group plus the identity matrix: 
\begin{equation}
	\rho=\sum \rho_\mu t_\mu,\hspace{0.25cm}H=\sum h_\mu t_\mu,\hspace{0.25cm}A_j=\sum a^j_\mu t_\mu,
	\end{equation} 
where $\mu$ is running from 0 to 8, being $t_0$ the identity matrix and $t_k$    the Gell-Mann matrices $(k=1,...,8)$, which satisfy $[t_a,t_b]=i\sum_c f_{abc}t_c$, where $f_{abc}$ are the structure constants of $SU(3)$. Imposing the increasing with time of the Von Neumman entropy the hermiticity of the 
$\hat{A}_j$ is assured, having, as a consequence, that the dissipative matrix can be expressed as~\cite{Gago02a}:
\begin{equation}    
D_{kj}=\frac{1}{2}\sum_{l,m,n}(a_{nl})f_{knm}f_{mlj}, \hspace{0.25cm}a_{nl}=\vec{a}_n.\vec{a}_l, 
\label{Dissipativematrix}
\end{equation}
being the matrix $\bf{D}\equiv$$ D_{kj}$ symmetric, with components $D_{\mu 0}=D_{0 \mu}=0$, and $\vec{a}_{r}=\{a^1_r, a^2_r, ...,a^8_r \}$. The complete positivity condition requires that the eigenvalues of the mixing matrix $\rho(t)$ should be positive at any time, this is achieved demanding that the matrix $\bf{A}\equiv$$ a_{nl}$ is positive~\cite{Benatti00,Benatti01}. The scalar product structure present in the elements $D_{kj}$ makes them to respect the Cauchy-Schwartz inequalities. Gathering the conservation of the probability to all that we have said, we have that the evolution equation of $\rho(t)$ is given by:
	\begin{equation}
	\label{Evolutionequation}
	\dot{\rho}_0=0,\hspace{0.5cm}\dot{\rho}_k=(H_{kj}+ D_{kj}) \rho_j =M_{kj}\rho_j,
	\end{equation}
	where $H_{kj}=\sum_i h_i f_{ijk}$. The solution of the  Eq.~(\ref{Evolutionequation}) written in matricial form is:
	\begin{equation}
	\label{SolEvoleq}
	\varrho (t)=e^{{\bf M}t}\varrho (0),
	\end{equation}
where $\varrho$ is an eight column vector compose by the $\rho_k$ and $\bf M$$\equiv M_{kj}$. Therefore, we can obtain a general expression for the neutrino oscillation probability
$\nu_\alpha\rightarrow\nu_\beta$:
	\begin{equation}
	\label{GeneralProbability}
	P_{\nu_\alpha\rightarrow\nu_\beta}=\frac{1}{3}+\frac{1}{2}(\varrho^{\beta})^T\varrho^{\alpha} (t) =  
\frac{1}{3}+\frac{1}{2}\sum_{i,j} \rho^{\beta}_{i}\rho^{\alpha}_{j}[e^{{\bf M}t}]_{ij} .
	\end{equation}	
Since in our analytical approach we will use the vacuum case, then, the 
$\rho_i^{\alpha}$ are already defined and they are given by:
	\begin{equation}
	\label{GeneralCoefficients}
	\begin{split}
	\rho_0^{\alpha}&=\sqrt{2/3},\\
	\rho_1^{\alpha}&=2\hspace{0.1cm}\mathrm{Re}\left(U^*_{\alpha1}U_{\alpha2}\right),\\
	\rho_2^{\alpha}&=-2\hspace{0.1cm}\mathrm{Im}\left(U^*_{\alpha1}U_{\alpha2}\right),\\
	\rho_3^{\alpha}&=|U_{\alpha1}|^2-|U_{\alpha2}|^2,\\
	\rho_4^{\alpha}&=2\hspace{0.1cm}\mathrm{Re}\left(U^*_{\alpha1}U_{\alpha3}\right),\\
	\rho_5^{\alpha}&=-2\hspace{0.1cm}\mathrm{Im}\left(U^*_{\alpha1}U_{\alpha3}\right),\\
	\rho_6^{\alpha}&=2\hspace{0.1cm}\mathrm{Re}\left(U^*_{\alpha2}U_{\alpha3}\right),\\
	\rho_7^{\alpha}&=-2\hspace{0.1cm}\mathrm{Im}\left(U^*_{\alpha2}U_{\alpha3}\right),\\
	\rho_8^{\alpha}&=\frac{1}{\sqrt{3}}\left(|U_{\alpha1}|^2+|U_{\alpha2}|^2-2|U_{\alpha3}|^2\right),\\
	\end{split}
	\end{equation}
where the $U_{\alpha j}$ refers to an element of the Pontecorvo-Maki-Nakagawa-Sakata(PMNS) \cite{Pontecorvo:1957cp,Maki:1962mu}.  If we want to solve the Eq. (\ref{GeneralProbability}) for  the antineutrino case is enough to make $U_{\alpha j}\rightarrow U^*_{\alpha j}$. 

\subsection{CPT violation and quantum decoherence}
\label{CPTdecoherenceviolation}

We will test the CPT symmetry in the context of DUNE using the simulated total rates associated to the $\nu_\mu$ and the $\bar{\nu}_\mu$ survival channels, where the matter effects are unimportant. The latter fact implies that the vaccum probabilities formulae for oscillation (plus decoherence) are going to be well enough for understanding the corresponding features of CPTV effects. Thus, all the formulae in this section will be developed under the vacuum framework. Before start, it is the utmost importance to remark that the decoherence phenomena entails the transition from pure to mixed states, which implies that the time reversal operation is, as itself, meaningless for this situation~\cite{Wald80}. The tool for revealing these, implicit, CPTV effects is the difference between the $\nu_\mu$ and $\bar{\nu}_\mu$ survival probabilities channels, which written for a generic flavor $\nu_\alpha$ is:  

	\begin{equation}
	\label{DefinitionAsymetryCPT}
	\Delta P_{\cancel{\mathrm{CPT}}}=	P_{\nu_\alpha \rightarrow \nu_\alpha}-P_{\bar{\nu}_\alpha \rightarrow \bar{\nu}_\alpha}.
	\end{equation}
With the aim of simplyfying of the analytical form of the latter expression we work under three assumptions: the diagonal elements (damping parameters) of the dissipative matrix ${\bf D}$ are all equal to a single parameter $\Gamma$, the dissipative matrix for neutrinos is equal to the corresponding for antineutrinos, ${\bf D}=\bar{{\bf D}}$, and the last is that the ${\bf D}$ matrix is containing no more than one non-diagonal elements at a time we study the $\Delta P_{\cancel{\mathrm{CPT}}}$. As a general feature, we have that a non-zero $\Delta P_{\cancel{\mathrm{CPT}}}$ is obtained when in the survival neutrino oscillation probability there is a term with $\beta_{ij}$ (non-diagonal term) coupled to $\rho_i^{\alpha}\rho_j^{\alpha}$ that contains $\sin \delta_{CP}$, therefore, when its corresponding antineutrino term is 
substracted for getting $\Delta P_{\cancel{\mathrm{CPT}}}$ they do not cancel each other because of the flipping of the sign 
of $\sin\delta_{CP}$. We find that the aforementioned situation (i.e. non null $\Delta P_{\cancel{\mathrm{CPT}}}$) is fulfilled by fifteen $\beta_{ij}$ where one coefficient in the product $\rho_i^{\alpha}\rho_j^{\alpha}$ is: $\rho_2^{\alpha}$, $\rho_5^{\alpha}$ or $\rho_7^{\alpha}$ and the other one : $\rho_1^{\alpha}$, $\rho_3^{\alpha}$, $\rho_4^{\alpha}$, $\rho_6^{\alpha}$ or $\rho_8^{\alpha}$ sumarizing in total fifteen cases. The remaining $\beta_{ij}$ does not produce non-null $\Delta P_{\cancel{\mathrm{CPT}}}$ given that 
they are not connected with $\rho_i^{\alpha}\rho_j^{\alpha}$ terms that contains $\sin \delta_{CP}$, similar to what happen for the survival probabilities, in the pure oscillation case, where there are no terms involving $\sin \delta_{CP}$ then these do not flip sign when we switch  
neutrinos to antineutrinos conserving CPT. 

Based on the similarities of the structure of the form for $\Delta P_{\cancel{\mathrm{CPT}}}$ we can divide these fifteen cases in two groups, each group related to different set of $\beta_{ij}$, that we present at follows.
 
\subsubsection{$\Delta P_{\cancel{\mathrm{CPT}}}$ for group one}
\label{Groupone}
The 
$\Delta P_{\cancel{\mathrm{CPT}}}$ expression for the first group is given by:
	\begin{equation}
	\label{AsymmetryCPTgroup1}
	\Delta P_{\cancel{\mathrm{CPT}}}=   \beta_{ij}\frac{\big(e^{\Omega_{\beta_{ij}}t}-e^{-\Omega_{\beta_{ij}}t}\big)}{\Omega_{\beta_{ij}}} \rho_i^\alpha \rho_j^\alpha \hspace{0.15cm}e^{-\Gamma t},
	\end{equation}
	where $\Omega_{\beta_{ij}}=\sqrt{{\beta_{ij}^2-\Delta_{\beta_{ij}}}^2}$, with $\Delta_{\beta_{ij}}=\Delta m_{\beta_{ij}}^2/2E$, where $E$ is energy and $\Delta m_{\beta_{ij}}^2$, corresponds to standard square mass differences of neutrino masses, according to its indexes $ij$ (see Table~\ref{TableGroup1}). This formula applies for nine $\beta_{ij}$, the details are given in Table~\ref{TableGroup1}. On the other hand, in Appendix~\ref{AppendixA}, as an example, we display in Eq.~(\ref{beta12})  the exact probability from where we can extrapolate the $\Delta P_{\cancel{\mathrm{CPT}}}$ for $\beta_{12}$.  
\begin{table}[!h]\begin{center}\begin{tabular}{c|c}
			$(i,j)$ & $\Delta_{\beta_{ij}}$ \\\hline
			$(1,2),(2,3),(2,8)$ & $\Delta_{12}$ \\
			$(4,5),(5,3),(5,8)$ & $\Delta_{13}$ \\
			$(6,7),(7,3),(7,8)$ & $\Delta_{23}$ \\
\end{tabular}
\caption{Here it is displayed each group of indexes $(i,j)$, which corresponds to a one of the nine $\beta_{ij}$. The $(i,j)$ in the same row are associated to the $\Delta_{\beta_{ij}}$ in the same line}
\label{TableGroup1}
\end{center}\end{table}

\subsubsection{$\Delta P_{\cancel{\mathrm{CPT}}}$ for group two}
\label{Grouptwo}
The $\Delta P_{\cancel{\mathrm{CPT}}}$ for the remaining six $\beta_{ij}$: $\beta_{15}, \beta_{24},\beta_{17}, \beta_{26}, \beta_{47}$ and $\beta_{56}$, is also proportional to $\beta_{ij}$, but it is rather a cumbersome expression in comparison to the one in Eq.~\ref{AsymmetryCPTgroup1}.  In fact, it is the addition of two terms, one of them is proportional to $\rho_{i}^\alpha \rho_{j}^\alpha$  while the other one, is proportional to $\rho_{k}^\alpha \rho_{l}^\alpha$. For a given $ij$ indexes, there is a specific $kl$, with each one of these indexes associated to an specific mass squared diference value, for the complete details see Table~\ref{TableConjugation}. The six expressions for the CPTV formula are obtained per each pair $ij,kl$ plus exchanging $ij  \leftrightarrow kl$, with all its correspondent terms associated with them. The explicit formula is given by:
	\begin{equation}
	\label{AsymmetryCPTgroup2}
	\begin{split}
	&\Delta P_{\cancel{\mathrm{CPT}}}=  \beta_{ij} \frac{1}{\sqrt{\Omega^4-4 \Delta_{\beta_{ij}}^2 \Delta_{\beta_{kl}}^2}}\\ &\times \bigg[ 
	\left(
	\Omega_{+}\big(e^{\Omega_{+}t}-e^{-\Omega_{+}t}\big)-\Omega_{-}\big(e^{\Omega_{-}t}-e^{-\Omega_{-}t}\big)
	\right)
	\rho_i^\alpha \rho_j^\alpha
	\\ &+\Delta_{\beta_{ij}} \Delta_{\beta_{kl}}
	\left(
	\frac{e^{\Omega_{+}t}-e^{-\Omega_{+}t}}{\Omega_{+}}-
	\frac{e^{\Omega_{-}t}-e^{-\Omega_{-}t}}{{\Omega_{-}}}
	\right)
	\rho_{k}^\alpha \rho_{l}^\alpha \bigg] e^{-\Gamma t},
	\end{split}
	\end{equation}
	where $\Omega^2=\beta_{ij}^2-\Delta_{\beta_{ij}}^2-\Delta_{\beta_{kl}}^2$
and $\Omega_{\pm}=\frac{1}{\sqrt{2}} \sqrt{\Omega^2\pm\sqrt{\Omega^4-4 \Delta_{\beta_{ij}}^2 \Delta_{\beta_{kl}}^2}}$. As in the case of group one, it is shown in Appendix~\ref{AppendixA} the probability for $\beta_{24}$ in Eq.~(\ref{beta24}). From there, the corresponding $\Delta P_{\cancel{\mathrm{CPT}}}$ can be extracted.

	\begin{table}[!h]\begin{center}\begin{tabular}{c}
			$\{{(i,j),\Delta_{\beta_{ij}}}\} \leftrightarrow \{{(k,l),\Delta_{\beta_{kl}}}\} $ \\\hline
			$ \{{(1,5), \Delta_{12}}\}\leftrightarrow \{{(2,4), \Delta_{13}}\}$ \\
			$\{{(1,7), \Delta_{12}}\} \leftrightarrow \{{(2,6), \Delta_{23}}\}$ \\
			$\{{(4,7), \Delta_{13}}\} \leftrightarrow \{{(5,6), \Delta_{23}}\} $\\
			\end{tabular}
			\caption{Here it is shown 
                how is the relation between the six indexes $(i,j)$ and $(k,l)$, each of  them associated to its corresponding $\beta$ and its neutrino mass square differences. }
			\label{TableConjugation}
	\end{center}\end{table}

\begin{figure}[!h]
	\centerline{
	\includegraphics[width=0.38\textwidth]{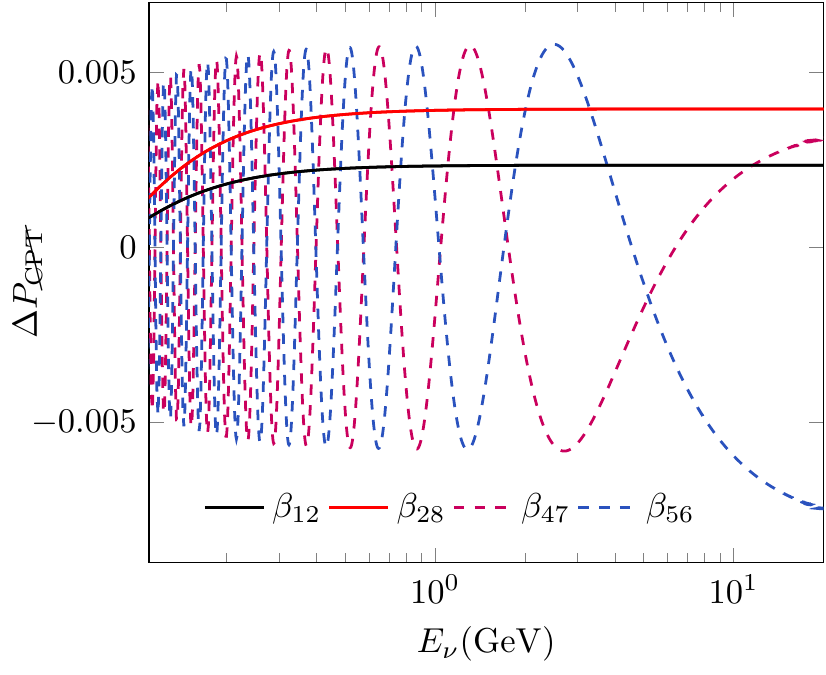}
	}		
	\caption{$\Delta P_{\cancel{\mathrm{CPT}}}$ versus 
		$E_\nu$, evaluated for $\Gamma = 10^{-23}$ GeV and $\delta_{CP}=3\pi/2$. This is for $\beta_{28}=\Gamma/\sqrt{3}$, $\beta_{12}=\Gamma/3$, $\beta_{56}=\Gamma/\sqrt{3}$ and $\beta_{47}=\Gamma/\sqrt{3}$. The remaining parameters are given 
		in Table~\ref{ParametersOscillation}.}
	\label{GraphCPTVvsEnergyVacuum}
\end{figure}
\begin{table}[!h]\begin{center}\begin{tabular}{c|c}
				Parameter & Value \\\hline
				$\theta_{12}$ & $33.63^{\circ}$ \\
				$\theta_{13} (\mathrm{NH})$ & $8.52^{\circ}$ \\
				$\theta_{23} (\mathrm{NH})$ & $48.7^{\circ}$ \\
				$\Delta m_{21}^{2} $ & $7.4 \times 10^{-5} \mathrm{eV}^{2}$ \\
				$\Delta m_{31}^{2}(\mathrm{NH}) $ & $2.515 \times 10^{-3} \mathrm{eV}^{2}$ \\
				Baseline & $1300 \mathrm{Km}$ \\
				\end{tabular}
			\caption{DUNE baseline and values for standard oscillation parameters taken from \cite{Nufit}.}
			\label{ParametersOscillation}
		\end{center}\end{table}
		
\subsubsection{$\Delta P_{\cancel{\mathrm{CPT}}}$ analytical results}
\label{analyticalACPT}

It is important to point out that, from now on, all the results that we will present the $\Delta P_{\cancel{\mathrm{CPT}}}$ wil be calculated for $\alpha=\mu$. In Fig. \ref{GraphCPTVvsEnergyVacuum},  we present $\Delta P_{\cancel{\mathrm{CPT}}}$ for a set of $\beta_{ij}$ per each group, which are: $\beta_{28}$, $\beta_{12}$ and 
$\beta_{47}$, $\beta_{56}$ for the group one and group two, respectively, and for neutrino energies from 0.1 to 20 GeV, which encloses the DUNE energy range. The selected $\beta$'s are those who produce a bigger amplitudes in the  $\Delta P_{\cancel{\mathrm{CPT}}}$. We have evaluated this effect in an isolated manner per each $\beta$ (i.e. considering all the rest of $\beta$'s as zero),  considering its maximum value which is obtained from the inequalities and positivity conditions given in Appendix \ref{Decoelementsconstraints}, having as result the following values: $|\beta_{28}|=\Gamma/\sqrt 3$, $|\beta_{12}|=\Gamma/3$ and  $|\beta_{47}|=|\beta_{56}|=\Gamma/\sqrt 3$, for these plots we have taken their positive values. For all these plots it is also fixed $\delta_{CP}=3\pi/2$ and $\Gamma =10^{-23}$ GeV, being that the remaining parameters are displayed in Table~\ref{ParametersOscillation}. The parameters given in 
Table~\ref{ParametersOscillation} will be used througout this paper. We note, for group one, that the $\beta_{28}$ is producing the highest amplitude for  $\Delta P_{\cancel{\mathrm{CPT}}}$, in all the energy range where a slightly minor effect for 
$\beta_{12}$ is observed. In the case of group two, $\beta_{47}$ and $\beta_{56}$ give the maximum values of amplitudes of $\Delta P_{\cancel{\mathrm{CPT}}}$ up to neutrino energies a bit less than 5 GeV.

\begin{figure}[h!]
	\centerline{
 		\includegraphics[scale=0.56]{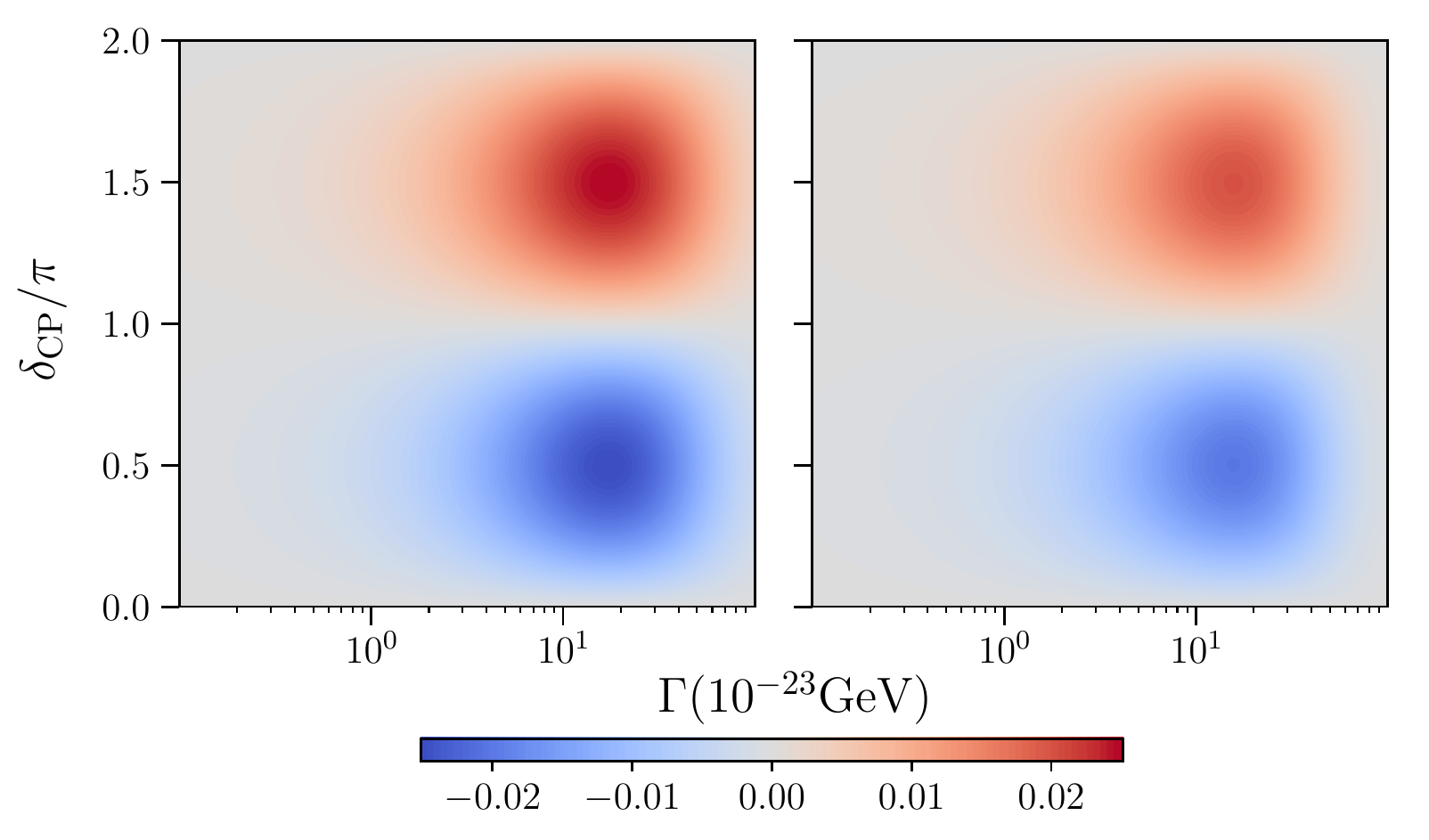}}	
 	\caption{Iso-contour curves of $\Delta P_{\cancel{\mathrm{CPT}}}$
 		at the plane $\Gamma$ versus $\delta_{CP}$ evaluated for $\beta_{28}=\Gamma/\sqrt{3}$ (left) and $\beta_{56}=\Gamma/\sqrt{3}$ (right) and 
 		for a fixed $E_\nu =2.4$ GeV.}
 	\label{GraphIsoVacuum}
 \end{figure}

In Fig. \ref{GraphIsoVacuum}, we have two plots 
which show iso-contour curves of $\Delta P_{\cancel{\mathrm{CPT}}}$
at the plane $\Gamma$ versus $\delta_{CP}$. For both plots the neutrino energy is fixed at 2.4 GeV keeping the remaining parameters at the same values than those used for Fig \ref{GraphCPTVvsEnergyVacuum}. One plot is for $\beta_{28}$ (group one) and the other for $\beta_{56}$ (group two), both taken equal to $\Gamma/\sqrt{3}$. As we said before these particular $\beta$'s are the ones who generate the biggest amplitudes for $\Delta P_{\cancel{\mathrm{CPT}}}$ per each group. Among the general features, we have that other than the maximum (and minimum) value of the $\Delta P_{\cancel{\mathrm{CPT}}}$ the behaviour of both plots is rather equal. Other common detail is that the $\Delta P_{\cancel{\mathrm{CPT}}}$ grows with $\Gamma$ until reaching a region where the maximum amplitude is located, then starts to decrease. Outside the regions around the peaks, i.e. for lower and higher values than the $\Gamma$ at the peak, the $\Delta P_{\cancel{\mathrm{CPT}}}$ is zero. For getting a full understanding of why happens this behaviour it is enough to look the formula given for group one, Eq.~(\ref{AsymmetryCPTgroup1}), since there is no a qualitative difference between the plots for $\beta_{28}$ (group one) and $\beta_{56}$ (group two). Hence, from Eq.~(\ref{AsymmetryCPTgroup1}), we see that $\Delta P_{\cancel{\mathrm{CPT}}}$ is suppressed for low values of $\Gamma$, which implies low values of $\beta_{28}(=\Gamma/\sqrt{3})$ that are directly proportional to the value of $\Delta P_{\cancel{\mathrm{CPT}}}$. On the other hand, $\Delta P_{\cancel{\mathrm{CPT}}}$ is reduced for higher values of $\Gamma$, given that the latter diminishes the factor $\exp^{-\Gamma t}$. From the maximization of the Eq.~(\ref{AsymmetryCPTgroup1}) the value of the $\Gamma$ at the peak can be extracted, for $\beta_{28}$ the peak is at $\Gamma \sim 1.7 \times 10^{-22}$ GeV, similarly, if we maximize the Eq.~(\ref{AsymmetryCPTgroup2}) we obtain the peak for $\beta_{56}$ at $\Gamma \sim 1.6 \times
10^{-22}$ GeV. In general, a very reasonable estimation for the value of $\Gamma$ at the  peak is obtained from $\Gamma L \sim 1$ then $\Gamma \sim 1/L$, which for $L=1300$ km is $\sim 1.5 \times 10^{-22}$ GeV. The corresponding values of $\delta_{CP}=\pi/2$ and $3\pi/2$, for $\beta_{28}$, can be directly inferred from the unique presence of $\sin \delta_{CP}$ in the factor $\rho_2^\mu \rho_8^\mu$, the values of $\delta_{CP}$ for $\beta_{56}$ are very close to $\pi/2$ and $3\pi/2$ for similar reasons, but they are slightly distorted due to $\rho_6^\mu$ is composed by two terms, being one of them is proportional to $\cos \delta_{CP}$. It is important to add that, in spite of they have been not showed here, we have checked that the equivalent plots of the Fig. \ref{GraphIsoVacuum}, when matter effects are included, do not reveal significant differences in comparison with the vacuum case presented here. Besides, of course, that given the presence of matter effects, a $\Delta P_{\cancel{\mathrm{CPT}}}\neq 0$ is expected even in the absence of decoherence. Actually, in the experimental (simulated) searches of CPTV that we will present in the following sections, the CPTV, due to matter effects, will play the role of normalization factor.       
 
	\begin{figure}[h!]
		\includegraphics[width=0.39\textwidth]{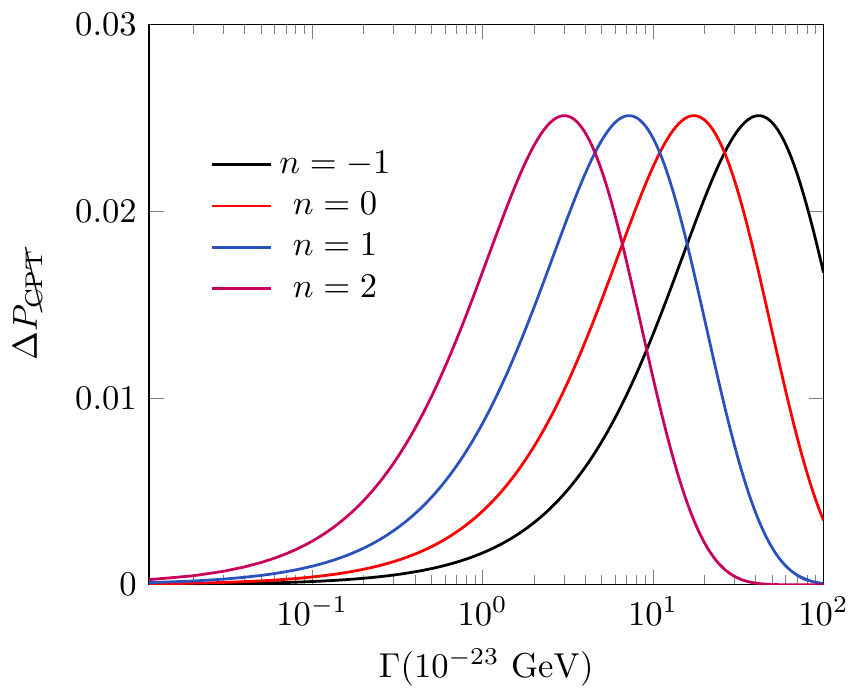}
		\caption{$\Delta P_{\cancel{\mathrm{CPT}}}$ versus 
			$\Gamma$, evaluated for different energy dependence
			$n = -1, 0, 1, 2$, $E_{\nu}=2.4$ GeV, $\delta_{CP}=3\pi/2$ 
			and fixing $\beta_{28}=\Gamma/\sqrt{3}$.}
		\label{PlotCPTVvsGamma}
	\end{figure}

\subsubsection{Decoherence parameters with energy dependency}
\label{Energydepend}	
From a more general view the entries of the decoherence matrix could be energy dependent, particularly, in this paper we will adopt this dependence as follows: 
	\begin{equation}
	\Gamma_{E_\nu}= \Gamma \left(\frac{E_\nu}{\mathrm{GeV}}\right)^{n}
	\label{EnergyDependeceGamma},
	\end{equation}	
where $n$ can be $-1,0,1$ and $2$. The $n=-1$ is taken because it imitates the oscillation energy dependence being that the motivation for $n=1$ and $n=2$ can be found in \cite{Liu97} and \cite{EllisMavro96}, respectively.

In Fig. \ref{PlotCPTVvsGamma} we study the $\Delta P_{\cancel{\mathrm{CPT}}}$  for the aforementioned energy dependence and setting $\beta_{28}=\Gamma/\sqrt{3}$, the neutrino energy in $2.4$ GeV (the DUNE energy peak) and $\delta_{CP}=3\pi/2$. In this figure we note that the energy dependency on $\Gamma$ only change its value at the peak but 
do not affect the amplitude of $\Delta P_{\cancel{\mathrm{CPT}}}$. As we have discussed in the section~\ref{analyticalACPT}, at the peak is satisfied approximately the next relation: $\Gamma_{E_\nu} L \sim 1$ then $\Gamma \sim 1/(L E^n$)  which turns out to 
be in $\Gamma \sim $ \{4.0, 1.5, 0.6, 0.3\}$ \times 10^{-22}$ GeV for 
$n = -1, 0, 1$ and $2$  respectively.  

\subsubsection{Optimal $\Delta P_{\cancel{\mathrm{CPT}}}$}
\label{optimalbetas}
For maximizing the $\Delta P_{\cancel{\mathrm{CPT}}}$ we simultaneously turn on  $\beta_{28}$, $\beta_{12}$, $\beta_{56}$ and $\beta_{47}$ in the following values: $\beta_{28}=\Gamma/\sqrt{3}$, $\beta_{12}=(\sqrt{2/3})\Gamma/3$ and  $\beta_{56}=-\beta_{47}= \Gamma/3$. These values has been set according to the following steps: First, we fix $\beta_{28}=\Gamma/\sqrt{3}$, given that this $\beta$  produces the major effect on $\Delta P_{\cancel{\mathrm{CPT}}}$. Second, once $\beta_{28}$ have been defined we obtain the maximum allowed value for $\beta_{12}$, which is the second in importance regarding to its impact on $\Delta P_{\cancel{\mathrm{CPT}}}$. By last, with $\beta_{28}$ and $\beta_{12}$ already set up, we get the maximum values of $\beta_{56}$ and $\beta_{47}$, where we have taking $\beta_{56}=-\beta_{47}$ in order to obtain a constructive effect between them. The restrictions imposed by the Schwarz inequalities and positivity conditions, fully described in the Appendix \ref{Decoelementsconstraints}, have been considered for getting the aforementioned values of $\beta$'s.
	
\subsubsection{CPT violation in matter}
\label{CPTVmatter}
As we have already mentioned, when the neutrinos are 
travelling through matter, we have a non-zero CPTV value 
for pure standard oscillation, even for zero CP phase. From now on, when we refer to the term standard oscillation (SO), it means that the matter effects
are included. If we add the decoherence to SO, the non-zero value of CPTV is still preserved, but, it has a different magnitude with respect to its corresponding in the pure SO, because, as expected, it is distorted by
the presence of the quantum decoherence parameters. In particular, it is interesting to note that this happens even when a single parameter diagonal decoherence matrix (DDM) (proportional to the identity) is considered. In constrast with the DDM case in vacuum, where a non-zero CPTV is not brought to light. The matter neutrino oscillation probabilities for a single parameter DDM can be derived only replacing the vacuum mixing angles and mass squared 
for their corresponding ones in matter, in, for instance, the three generation formula displayed in~\cite{Gago02a}. Of course, it also includes the replacement of a singular decoherence parameter. The application of the latter procedure is fully justified and it has been very well explained in~\cite{Carpio:2017nui}. Therefore, we have that the structure of the formula is given by:       
\begin{equation}
P^{\mathrm{SO \bigoplus DDM}}_{\nu_{\alpha}\nu_{\rho}}
=\frac{1}{3}(1-e^{-\Gamma t}) +
e^{-\Gamma t} P^{\mathrm{SO}}_{\nu_{\alpha}\nu_{\rho}},
\label{pdiag}
\end{equation}
where $\alpha$, $\rho$, are neutrino flavours, and $\mathrm{SO} \ (\mathrm{SO \bigoplus DDM})$ stands for standard oscillation (standard oscillation plus diagonal decoherence). It is clear that: $\Delta P^{\mathrm{SO \bigoplus DDM}}_{\cancel{\mathrm{CPT}}}= e^{-\Gamma t} \Delta P^{\mathrm{SO}}_{\cancel{\mathrm{CPT}}}$, which goes to zero for 
high values of $\Gamma$. Nonetheless, when we deal with a real situation, the latter does not occurs, since, we have to convolute the neutrino (antineutrino) oscillation probabilities with the neutrino (antineutrino) fluxes, cross sections, efficiencies and resolution, being that, for this context, the 1/3 from the first term at the RHS in Eq.(\ref{pdiag}) is the only that survives for high values of $\Gamma$, leading us to find a non-zero constant value. We will see this type of behaviour further ahead in our section of results. 
   
In this paper we are not going derive analytical formula for the neutrino matter oscillation probability, for the non-diagonal decoherence matrices (NDM) cases that we have presented before. This is because it is a rather complicated task and, besides, as we have already argued, the vacuum oscillation probabilities formulas are going to be enough for having a qualitative understanding of our results.

\section{Experiment, Simulation and Results}

The DUNE experiment will be able to unravel several non-standard neutrino physics scenarios through oscillation measurements \cite{marvin,MasudDUNE,Berryman}. It will consist in a muon neutrino(antineutrino) beam traversing the Earth from Fermilab to Sanford Underground Research Facility (SURF) which comprises a distance of 1300 km and  average matter density of $\rho_{\mathrm{DUNE}}=2.96 \ \mathrm{g}/\mathrm{cm}^3$. At SURF the neutrino beam will hit a massive liquid argon time-projection chamber (LArTPC) of 40 Ktons~\cite{cdr}.


For this work, it is assumed the configuration of $80$ $\mathrm{GeV}$ energy with $1.07	$ $\mathrm{MW}$ power in the primary proton beam from the Main Injector runing over 5 years for exposure for each mode (FHC and RHC). In our simulation of DUNE, the  GLoBES package~\cite{globes1,globes2} is used and feeding with the information of the cross section, neutrino fluxes, resolution function and efficiency extracted from \cite{ancillaryfiles}. While, the matter neutrino oscillation probabilities plus decoherence was calculated with nuSQuIDS \cite{nusquids}. 

	\begin{figure}[h!]
		\centerline{
			\includegraphics[scale=0.56]{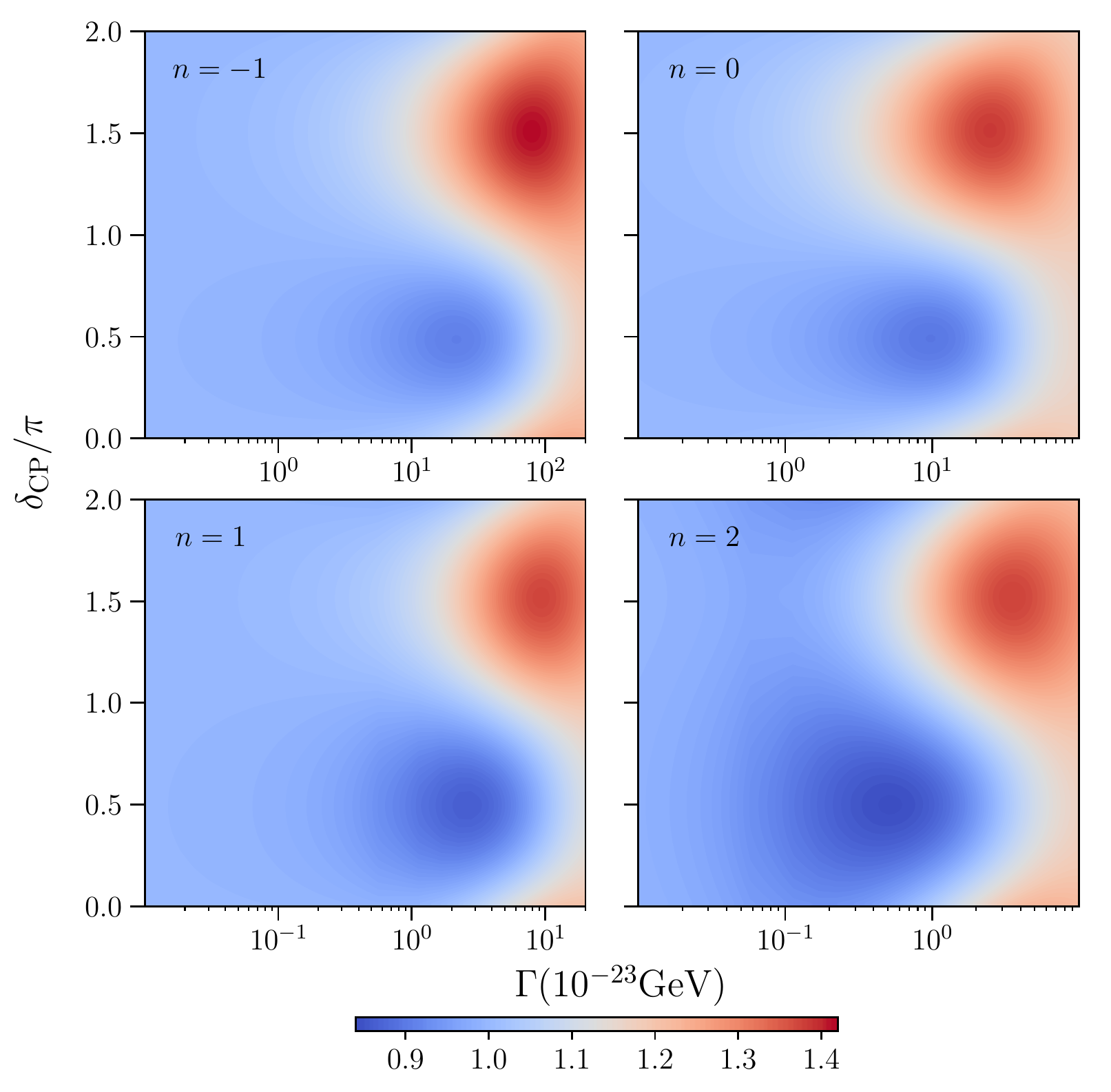}    
		}
		
		\caption{We analize the confidence levels for the maximum values for $\mathcal{R}$. For $\delta_{\mathrm{CP}}\sim \pi/2$ we have $2.9  \sigma$, $3.4  \sigma$, $4.7  \sigma$ and $5.5  \sigma$ of confidence for $n = -1$, 0, 1 and 2 respectively. On the other hand, for $\delta_{\mathrm{CP}}\sim 3\pi/2$ we have  $10.3  \sigma$, $9.8  \sigma$, $9.6  \sigma$ and $9.7  \sigma$ of confidence for $n = -1$, 0, 1 and 2 respectively.}
		\label{PlotIsoclinesResults}
	\end{figure}

For testing the CPTV effects the following experimental 
observable is defined: 
\begin{equation}
\mathcal{R}= \frac{\Delta N^{\mathrm{SO \bigoplus DEC}}}{ \Delta N^{\mathrm{SO}}},
\label{ObservableR}
\end{equation}
where $\Delta N^{\mathrm{SO} \ (\mathrm{SO \bigoplus DEC})} = N_{\nu_\mu} - N_{\bar{\nu}_\mu}$ is the difference between the total events rates for neutrino and antineutrino, respectively, and DEC stands for any case of decoherence.

The total event rates has been 
calculated using the prescription given in \cite{cdr}. Our observable 
is normalized with the SO difference of events  
$\Delta N^{\mathrm{SO}}$, which is non-zero due to 
matter effects plus the intrinsic differences between the 
 cross sections, fluxes, etc for neutrinos and 
antineutrinos. Given our definition, when decoherence is absent $\mathcal{R}=1$.

For giving an idea of the impact 
of decoherence into SO
physics, we display 
in Table \ref{TableEvents}, the total rates for four 
energy dependent decoherence scenarios.

In Fig. \ref{PlotIsoclinesResults} we are showing iso-contour curves for the 
observable $\mathcal{R}$ in the plane $\Gamma$ versus $\delta$  for four plots  which corresponds to $n=-1,0,1,$ and $2$. In these plots, the maximum amplitudes are located at similar $\delta_{CP}$, $\delta_{CP} \simeq \pi/2$ and $3\pi/2$, to those presented in Fig. \ref{GraphIsoVacuum}. In relation to the Fig. \ref{GraphIsoVacuum}, there is a dislocation between the values of $\Gamma$ at the maximum amplitudes for $\delta_{CP} \simeq \pi/2$ and $3\pi/2$. This is mainly because of the differences in the inputs used when we convolute the probabilities for the neutrino and antineutrino mode. In addition, the $\Gamma$ for $\delta_{CP}\simeq \pi/2$ and $3\pi/2$  is shifted to its lower values whenever $n$ increases, gaining more sensitivity to lower values of $\Gamma$. The latter kind of behaviour is expected and resembles the one we have seen for $\Delta P_{\cancel{\mathrm{CPT}}}$ in Fig. \ref{PlotCPTVvsGamma} (but here is a one dimensional view). Moreover, 
we also see the existance of degeneracies in ($\Gamma, \delta$) likewise we have in \cite{Carpio:2017nui}. 



\begin{figure}[h!]
		\vspace{-4pt}
		\centerline{
			\includegraphics[width=0.48\textwidth]{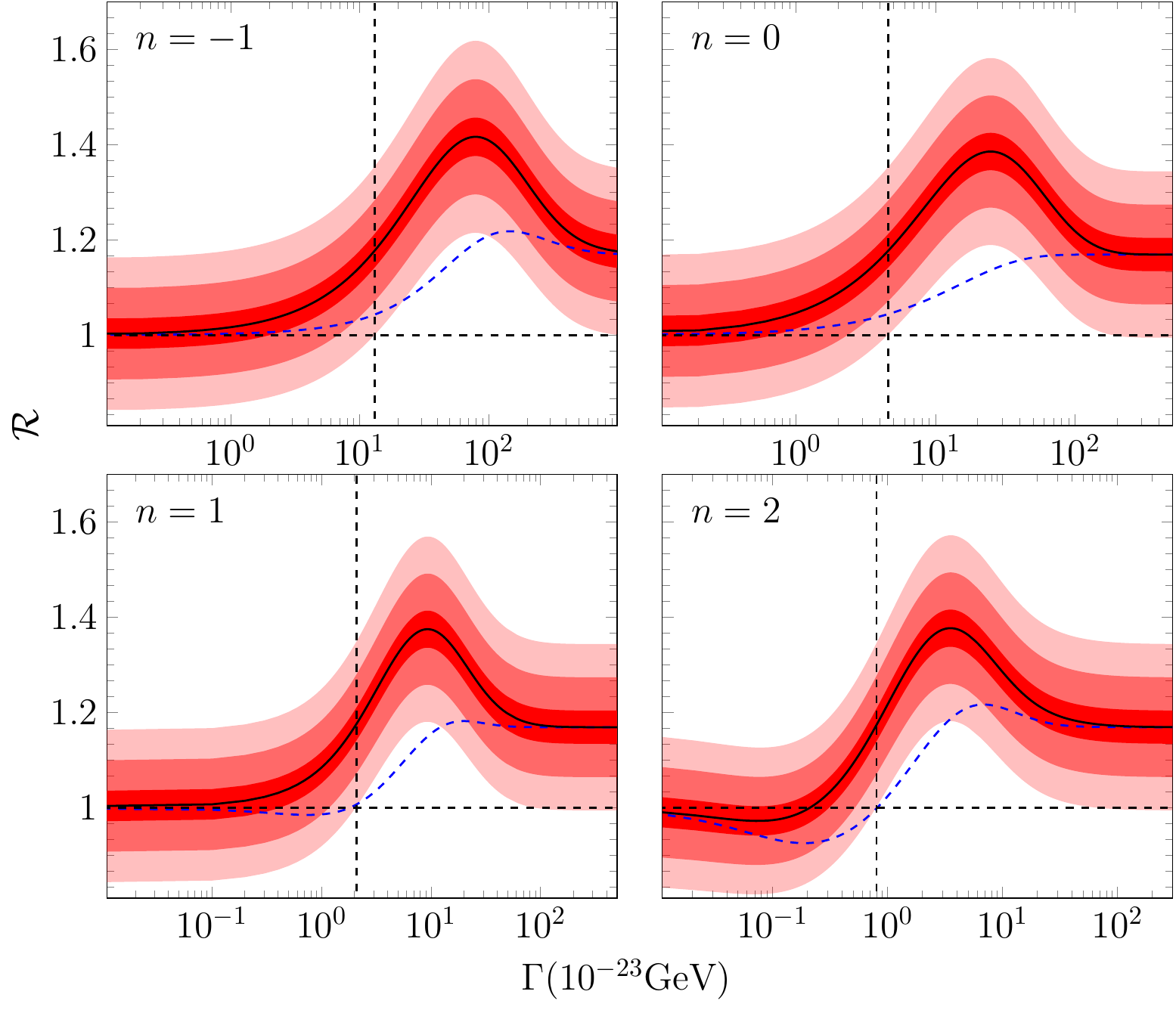}
		}
		\caption{The black horizontal dashed line is the expected in the SO. The blue dashed line corresponds to the case of a DDM. Meanwhile, the solid black line corresponds to the case of a NDM both cases evaluated at $\delta_{CP} = 3\pi/2$. The red fringes (small, medium and large) represent the statistical error $1\sigma$, $3\sigma$ and $5\sigma$ (respectively). The $\beta$'s used corresponds to the ones given at 
section~\ref{optimalbetas}. The intersection between the black horizontal dashed line with the vertical one marks the $5\sigma$ significance of the NDM case relative to the SO case.}
		\label{PlotErrorBarResults3pi2}
	\end{figure}

In Fig. \ref{PlotErrorBarResults3pi2}, we present the observable  $\mathcal{R}$, with its corresponding error bands for $1\sigma$, $3\sigma$ and $5\sigma$, versus $\Gamma$, for $n=-1,0,1,$ and $2$. We take $\delta_{CP} = 3\pi/2$
given that we learn from Fig. \ref{PlotIsoclinesResults} that one of the maximum amplitude of $\Delta P_{\cancel{\mathrm{CPT}}}$ is obtained  
at this $\delta_{CP}$. The behaviour displayed in this plot for small and medium values of $\Gamma$, at the given scale,  is rather similar than that observed 
in Fig. \ref{PlotCPTVvsGamma}. However, for large values of $\Gamma$ the observable
$\mathcal{R} \sim 1.17$, and not  $\sim 1.0$, which it could be expected taking into 
account only the signal. This discrepancy is due to the inclusion of the backgrounds in our calculations, as described at Table \ref{TableEvents}. In order to make a comparison, we introduce in this plot the $\mathcal{R}$ corresponding to the single parameter DDM. We see that at small and large values of $\Gamma$, $\mathcal{R}$ tends to be $\sim 1.$ and $\sim 1.17$, respectively, for the DDM and NDM, irregardless the dependency on $n$, as well. As we have anticipated in section \ref{CPTVmatter}, the diagonal case also produces  non-zero $\Delta P_{\cancel{\mathrm{CPT}}}$ but in a lower magnitude than the NDM case. In fact, we have that for the NDM case a $5\sigma$ discrepancy, respect to the expectation value for SO ( $\mathcal{R}=1$), is reached at the following $\Gamma =$ $\{13.1, 4.6, 2.1,0.8\} \times 10^{-23} \mathrm{GeV}$  for $n=-1,0,1$ and $2$, respectively. It is interesting to note that at these values of $\Gamma$ the
DDM is compatible with the SO prediction. Thus, here, we would be able to distinguish the NDM from the DDM. 


	\begin{table}[h!]
        \begin{center}\begin{tabular}{|c c c c c c|}
		\hline
		\ $\Gamma = 10^{-23}\hspace{0.2cm}\mathrm{GeV}$ & Std & $n=-1$ & $n=0$ & $n=1$ & $n=2$ \\
		\hline
		$\mathbf{Neutrino} \ \mathbf{mode} $ & & & & & \\
		$\nu_{\mu} \ \mathrm{Signal}$ & $ 11025 $ & $11120$ & $11263$ & $11017$ &  $11524$\\
		$\bar{\nu}_{\mu} \ \mathrm{CC} \ \mathrm{Background}$  & $724$ & $721$ & $702$ & $556$ & $408$\\
		$\mathrm{NC} \ \mathrm{Background}$ & $109$ & $109$ & $109$ & $109$ & $109$\\
		$\nu_{\tau} + \bar{\nu}_{\tau} \ \mathrm{CC} \ \mathrm{Background}$ & $43$ & $43$ & $46$ & $74$ & $87$\\
		\hline
		$\mathbf{Antineutrino} \ \mathbf{mode} $ & & & & & \\
		$\bar{\nu}_{\mu} \ \mathrm{Signal}$ & $3754$ & $3752$ & $3749$ & $3557$ & $3555$\\
		$ \nu_{\mu} \ \mathrm{CC} \ \mathrm{Background}$ & $2149$ & $2145$ & $2097$ & $1680$ & $1261$\\
		$\mathrm{NC} \ \mathrm{Background}$ & $58$ & $58$ & $58$ & $58$ & $58$\\
		$\nu_{\tau} + \bar{\nu}_{\tau} \ \mathrm{CC} \ \mathrm{Background} $ & $27$ & $27$ & $29$ & $50$ & $60$ \\
		\hline
		\end{tabular}
		\caption{Total rates for the signal of $\nu_\mu$ and $\bar{\nu}_\mu$ disappearance channels and their corresponding background.}
		\label{TableEvents}
	\end{center}\end{table}

\begin{figure}[h!]
		\vspace{-4pt}
		\centerline{
			\includegraphics[width=0.48\textwidth]{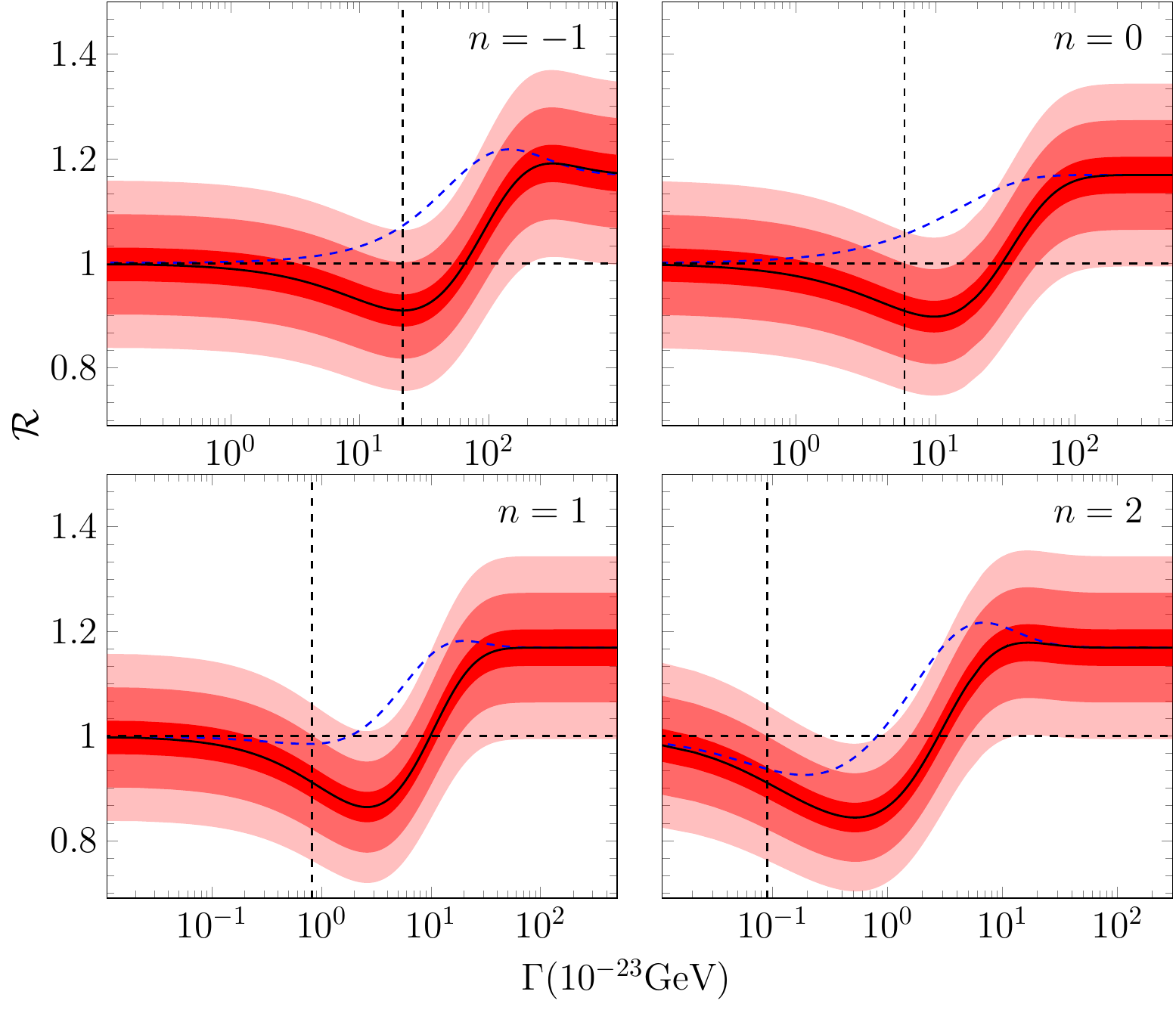}
		}
		\caption{Similar to Fig.~\ref{PlotErrorBarResults3pi2}, but for $\delta_{\mathrm{CP}} = \pi /2$. Here, the intersection between the black horizontal dashed line with the vertical one marks the $3\sigma$ significance of the NDM case relative to the SO case.}
		\label{PlotErrorBarResultspi2}
	\end{figure}

An analogous result is shown n Fig. \ref{PlotErrorBarResultspi2}, but taking $\delta_{CP} = \pi/2$. In this case, the following values of $\Gamma$ achieve the 3$\sigma$ significance:      	
$\{21.6, 6, 0.8,0.09\} \times 10^{-23} \mathrm{GeV}$  for $n=-1,0,1$ and $2$, respectively. 
All of them have $\mathcal{R} <1 $  For the cases $n=-1,0$, we can discriminate between the NDM and DDM, since we have: $\mathcal{R} < 1$ and $\mathcal{R} > 1$, respectively. For $n=1$, the DDM case is congruent with the SO, meanwhile, for $n=2$, the DDM and NDM can be confused.  
 

	\section{SUMMARY AND CONCLUSIONS}

We have shown that an apparent breakdown of the fundamental CPT symmetry can take place when the neutrino system is affected by the environment. This CPTV is produces by the combination of having $\delta_{\mathrm{CP}}$ in the neutrino sector with a certain set of some non-null coherences terms in the dissipative matrix. Furthermore, we have quantified a possible measurement of this CPTV using 
the dissapearance channels $\nu_{\mu} \rightarrow \nu_{\mu}$ and $\bar{\nu}_{\mu} \rightarrow \bar{\nu}_{\mu}$, with their corresponding backgrounds,
and an observable $\mathcal{R}$. All in the 
context of the DUNE experiment. The simulated measurements 
of $\mathcal{R}$ have been performed considering four hypothesis of energy dependence on the decoherence parameters: $n=-1,0,1,$ and $2$, where $\Gamma_{E_\nu}=\Gamma (E_\nu/\mathrm{GeV})^n$. For $\delta_{CP}=3\pi/2$, which is rather close to the current value of $\delta_{CP}$ given by the global fit~\cite{Nufit}, and a NDM, we achieve a $5 \sigma$ 
for $\mathcal{R}$ with respect to its expectation value at the SO case, $\mathcal{R}=1$, for the following $\Gamma$: $\{13.1, 4.6, 2.1,0.8\} \times 10^{-23} \mathrm{GeV}$, for $n=-1,0,1$ and $2$, respectively. At all these points, the DDM is compatible with the SO case. For $\delta_{CP}=\pi/2$,
we reach discrepancies of the order of $3 \sigma$. In our best case for $n=2$ we have
$\Gamma \simeq 10^{-24} \mathrm{GeV}$, but with the inability of discriminating from the DDM case.  We have to 
keep on mind that the aforementioned observations of CPTV appear 
when the neutrino system is treated as an open system. The latter means that it is likely that if we had access to the information of the environment, i.e. to the whole system, the overall CPT symmetry is conserved. For this reason, it deserves a more profound discussion to ascertain if this CPTV is a breaking at the fundamental level or it is only an apparent one, because of our lack of information from the environment. In some way, this CPTV represents a loss of information that in order to show that this information is not destroyed we need to know how this CPTV is compensated with the environment, probing the 
conservation of the information.    
	
	\section{ACKNOWLEDGEMENTS}
	AMG acknowledges funding by the {\it Direcci\'on de Gesti\'on de la Investigaci\'on} at PUCP, through grant  DGI-2017-3-0019. JCC and FND acknowledge CONCYTEC for the two graduate  fellowship  under  Grant  No. 233-2015-2-CONCYTEC-FONDECYT and  000236-2015-FONDECYT-DE.  We  also  thank S. Hern\'andez, E. Massoni and J. Hoefken for useful discussions.
	
	\appendix
	\section{Some probabilities formulae}
	\label{AppendixA}
	
	The dissipative matrix defined by Eq. (\ref{Dissipativematrix}) can be parameterized, in general, by 36 free parameters in the following form,

	\begin{equation}
	\bf{D}=\begin{pmatrix}
	-\gamma_1&\beta_{12}&\beta_{13}&\beta_{14}&\beta_{15}&\beta_{16}&\beta_{17}&\beta_{18}\\
	\beta_{12}&-\gamma_2&\beta_{23}&\beta_{24}&\beta_{25}&\beta_{26}&\beta_{27}&\beta_{28}\\
	\beta_{13}&\beta_{23}&-\gamma_3&\beta_{34}&\beta_{35}&\beta_{36}&\beta_{37}&\beta_{38}\\
	\beta_{14}&\beta_{24}&\beta_{34}&-\gamma_4&\beta_{45}&\beta_{46}&\beta_{47}&\beta_{48}\\
	\beta_{15}&\beta_{25}&\beta_{35}&\beta_{45}&-\gamma_5&\beta_{56}&\beta_{57}&\beta_{58}\\
	\beta_{16}&\beta_{26}&\beta_{36}&\beta_{46}&\beta_{56}&-\gamma_6&\beta_{67}&\beta_{68}\\
	\beta_{17}&\beta_{27}&\beta_{37}&\beta_{47}&\beta_{57}&\beta_{67}&-\gamma_7&\beta_{78}\\
	\beta_{18}&\beta_{28}&\beta_{38}&\beta_{48}&\beta_{58}&\beta_{68}&\beta_{78}&-\gamma_8\\
	\end{pmatrix}.
	\end{equation}	
			
	However, we will focus on the survival probabilities for only two cases: $\beta_{12}$ and $\beta_{24}$, below we display these probabilities.
	
For $\beta_{12}$:
	\begin{equation}
	\begin{split}
		&P_{\nu_{\alpha}\nu_{\alpha}}=\frac{1}{3}+\frac{1}{2}\bigg(\big((\rho^{\alpha}_{1})^2+(\rho^{\alpha}_{2})^2\big)\frac{\big(e^{\Omega_{12}t}+e^{-\Omega_{12}t}\big)}{2}\\
		&+\big( (\rho^{\alpha}_{4})^2+(\rho^{\alpha}_{5})^2
		\big)\cos{\Delta_{13} t}+\big((\rho^{\alpha}_{6}\big)^2+(\rho^{\alpha}_{7})^2\big)\cos{\Delta_{23} t}\\
		&+(\rho^{\alpha}_3)^2+(\rho^{\alpha }_8)^2+\beta_{12}\frac{\big(e^{\Omega_{12}t}-e^{-\Omega_{12}t}\big)}{\Omega_{12}}\rho_1^{\alpha}\rho_2^{\alpha}\bigg)e^{-\Gamma t}.\\
	\end{split}
		\label{beta12} 
	\end{equation}
	
		For $\beta_{24}$:
	\begin{equation}
	\begin{split}
		&P_{\nu_{\alpha}\nu_{\alpha}}=\frac{1}{3}+\frac{e^{-\Gamma t}}{2}\bigg(\frac{\big(e^{\Omega_{+}t}+e^{-\Omega_{+}t}\big)}{2\sqrt{\Omega^4-4\Delta_{12}^2\Delta_{13}^2}}\times g^{-, 1}_{+, 2}\\
		&+\frac{\big(e^{\Omega_{-}t}+e^{-\Omega_{-}t}\big)}{2\sqrt{\Omega^4-4\Delta_{12}^2\Delta_{13}^2}}g^{+,1}_{-,2}+\frac{\big(e^{\Omega_{+}t}+e^{-\Omega_{+}t}\big)}{2\sqrt{\Omega^4-4\Delta_{12}^2\Delta_{13}^2}}g^{+,4}_{-,5}\\
		&+\frac{\big(e^{\Omega_{-}t}+e^{-\Omega_{-}t}\big)}{2\sqrt{\Omega^4-4\Delta_{12}^2\Delta_{13}^2}}g^{-,4}_{+,5}+\big((\rho^{\alpha}_{6}\big)^2+(\rho^{\alpha}_{7})^2\big)\cos{\Delta_{23} t}\\
		&+(\rho^{\alpha}_3)^2+(\rho^{\alpha }_8)^2+\beta_{24} \frac{1}{\sqrt{\Omega^4-4 \Delta_{12}^2 \Delta_{13}^2}}\\ &\times \bigg[ 
		\left(
		\Omega_{+}\big(e^{\Omega_{+}t}-e^{-\Omega_{+}t}\big)-\Omega_{-}\big(e^{\Omega_{-}t}-e^{-\Omega_{-}t}\big)
		\right)
		\rho_2^\alpha \rho_4^\alpha
		\\
		&+\Delta_{12} \Delta_{13}
		\left(
		\frac{e^{\Omega_{+}t}-e^{-\Omega_{+}t}}{\Omega_{+}}-
		\frac{e^{\Omega_{-}t}-e^{-\Omega_{-}t}}{{\Omega_{-}}}
		\right)
		\rho_{1}^\alpha \rho_{5}^\alpha \bigg]\bigg),
	\end{split}
		\label{beta24} 
	\end{equation}
	where: 
\begin{equation*}
\begin{split}
g^{(\pm)_{\text{up}}, i}_{(\pm)_{\text{down}}, j}=\big( (\pm)_{\text{up}}(\Delta_{12}^2+{\Omega'}_{(\pm)_{\text{up}}}^2)(\rho^{\alpha}_{i})^2 \\ (\pm)_{\text{down}}(\Delta_{13}^2+{\Omega'}_{(\pm)_{\text{down}}}^2)(\rho^{\alpha}_{j})^2\big),
\end{split}
\end{equation*}
with
	${\Omega'}_{\pm}^2= \frac{1}{2} \big( \Omega^2\pm\sqrt{\Omega^4-4 \Delta_{12}^2 \Delta_{13}^2}\big)$.
	
	\section{Constraints for the decoherence matrix elements}
	\label{Decoelementsconstraints}
	For the two-flavor and three-flavor case, the conditions for the decoherence entries can be found in~\cite{Benatti01} and~\cite{Hernandez16}, respectively. Here we display the latter: 	
\begin{equation*}
	0\leq   | \Vec{a}_1   |=-\gamma_1+\gamma_2+\gamma_3-\frac{1}{3}\gamma_8,
\end{equation*}
\vspace*{-0.8cm}	
\begin{equation*}
0\leq   | \Vec{a}_2 |=\gamma_1-\gamma_2+\gamma_3-\frac{1}{3}\gamma_8,
\end{equation*}	
\vspace*{-0.8cm}	
\begin{equation*}
0\leq   | \Vec{a}_3 |=\gamma_1+\gamma_2-\gamma_3-\frac{1}{3}\gamma_8,
\end{equation*}
\vspace*{-0.8cm}		
\begin{equation*}
	0\leq   | \Vec{a}_4 |=-\gamma_4+\gamma_5+
	\frac{2}{3}\gamma_8-\frac{2}{\sqrt{3}}\beta_{38},
\end{equation*}
\vspace*{-0.8cm}		
\begin{equation*}
	0\leq   | \Vec{a}_5 |=\gamma_4-\gamma_5+
	\frac{2}{3}\gamma_8-\frac{2}{\sqrt{3}}\beta_{38},
\end{equation*}
\vspace*{-0.8cm}		
\begin{equation*}
	0\leq   | \Vec{a}_6 |=-\gamma_6+\gamma_7+
	\frac{2}{3}\gamma_8+\frac{2}{\sqrt{3}}\beta_{38},
\end{equation*}
\vspace*{-0.8cm}		
\begin{equation*}
0\leq   | \Vec{a}_7 |=\gamma_6-\gamma_7+
	\frac{2}{3}\gamma_8+\frac{2}{\sqrt{3}}\beta_{38},
\end{equation*}
\vspace*{-0.8cm}		
\begin{equation}
\begin{split}
0&\leq   | \Vec{a}_8 |=-\frac{1}{3}\gamma_1-\frac{1}{3}\gamma_2-\frac{1}{3}\gamma_3\\&\hspace{0.5cm}+\frac{2}{3}\gamma_4+\frac{2}{3}\gamma_5+\frac{2}{3}\gamma_6+\frac{2}{3}\gamma_7-\gamma_8.
\end{split}
\end{equation}	
	
Being their Schwartz inequalities: 
	
	\begin{equation*}
	4 {\beta_{12}}^2 \leq  \left(\gamma_3-\frac{\gamma_8}{3} \right)^2-\left(\gamma_1-\gamma_2 \right)^2,
	\end{equation*}
\vspace*{-0.8cm}	
	\begin{equation*}
	4{\beta_{13}}^2 \leq  \left(\gamma_2-\frac{\gamma_8}{3} \right)^2-\left(\gamma_1-\gamma_3 \right)^2,
	\end{equation*}
\vspace*{-0.8cm}	
	\begin{equation*}
	{4}{\beta_{23}}^2 \leq   \left(\gamma_1-\frac{\gamma_8}{3} \right)^2-\left(\gamma_2-\gamma_3 \right)^2,
	\end{equation*}
\vspace*{-0.8cm}	
	\begin{equation*}
	4{\beta_{45}}^2 \leq   \left(\frac{2\gamma_8}{3}-\frac{2\beta_{38}}{\sqrt{3}} \right)^2-\left(\gamma_4-\gamma_5 \right)^2,	
	\end{equation*}
\vspace*{-0.8cm}	
	\begin{equation*}
	4{\beta_{67}}^2 \leq  \left(\frac{2\gamma_8}{3}+\frac{2\beta_{38}}{\sqrt{3}} \right)^2-\left(\gamma_6-\gamma_7 \right)^2,
	\end{equation*}

\vspace*{-0.8cm}

	\begin{equation*}
	{\left(\frac{2}{3}\beta_{38}+\frac{1}{\sqrt{3}}\gamma_4+\frac{1}{\sqrt{3}}\gamma_5-\frac{1}{\sqrt{3}}\gamma_6-\frac{1}{\sqrt{3}}\gamma_7 \right)}^2 \leq   | \Vec{a}_3 | | \Vec{a}_8 |,	
	\end{equation*}
\vspace*{-0.8cm}	
	\begin{equation*}
	{\left(\frac{1}{\sqrt{3}}\beta_{16}-\frac{1}{\sqrt{3}}\beta_{27}+\frac{1}{\sqrt{3}}\beta_{34}+\frac{5}{3}\beta_{48} \right)}^2 \leq   | \Vec{a}_4 | | \Vec{a}_8 |,	
	\end{equation*}
\vspace*{-0.8cm}	
	\begin{equation*}
	{\left(\frac{1}{\sqrt{3}}\beta_{17}+\frac{1}{\sqrt{3}}\beta_{26}+\frac{1}{\sqrt{3}}\beta_{35}+\frac{5}{3}\beta_{58} \right)}^2  \leq   | \Vec{a}_5 | | \Vec{a}_8 |,	
	\end{equation*}
\vspace*{-0.8cm}	
	\begin{equation*}
	{\left(\frac{1}{\sqrt{3}}\beta_{14}+\frac{1}{\sqrt{3}}\beta_{25}-\frac{1}{\sqrt{3}}\beta_{36}+\frac{5}{3}\beta_{68} \right)}^2  \leq   | \Vec{a}_6 | | \Vec{a}_8 |,	
	\end{equation*}
\vspace*{-0.8cm}		
	\begin{equation*}
	{\left(\frac{1}{\sqrt{3}}\beta_{15}-\frac{1}{\sqrt{3}}\beta_{24}-\frac{1}{\sqrt{3}}\beta_{37}+\frac{5}{3}\beta_{78} \right)}^2  \leq   | \Vec{a}_7 | | \Vec{a}_8 |,	
	\end{equation*}
\vspace*{-0.8cm}		
	\begin{equation*}
	{\left(\beta_{14}-\beta_{25}+\beta_{36}+\frac{1}{\sqrt{3}}\beta_{68}  \right)}^2 \leq   | \Vec{a}_1 | | \Vec{a}_4 |,	
	\end{equation*}
\vspace*{-0.8cm}	
	\begin{equation*}
	{\left(\beta_{15}+\beta_{24}+\beta_{37}+\frac{1}{\sqrt{3}}\beta_{78}  \right)}^2 \leq   | \Vec{a}_1 | | \Vec{a}_5 |,	
	\end{equation*}
\vspace*{-0.8cm}	
	\begin{equation*}
	{\left(\beta_{16}+\beta_{27}-\beta_{34}+\frac{1}{\sqrt{3}}\beta_{48}  \right)}^2 \leq   | \Vec{a}_1 | | \Vec{a}_6 |,	
	\end{equation*}
\vspace*{-0.8cm}	
	\begin{equation*}
	{\left(\beta_{17}-\beta_{26}-\beta_{35}+\frac{1}{\sqrt{3}}\beta_{58}  \right)}^2 \leq   | \Vec{a}_1 | | \Vec{a}_7 |,
	\end{equation*}
\vspace*{-0.8cm}	
	\begin{equation*}
	{\left(\frac{2}{3}\beta_{18}-\frac{2}{\sqrt{3}}\beta_{46}-\frac{2}{\sqrt{3}}\beta_{57}  \right)}^2 \leq   | \Vec{a}_1 | | \Vec{a}_8 |,
	\end{equation*}
\vspace*{-0.8cm}	
	\begin{equation*}
	{\left(\beta_{15}+\beta_{24}-\beta_{37}-\frac{1}{\sqrt{3}}\beta_{78}  \right)}^2 \leq   | \Vec{a}_2 | | \Vec{a}_4 |,
	\end{equation*}
\vspace*{-0.8cm}	
	\begin{equation*}
	{\left(\beta_{14}-\beta_{25}-\beta_{36}-\frac{1}{\sqrt{3}}\beta_{68}  \right)}^2 \leq   | \Vec{a}_2 | | \Vec{a}_5 |,	
	\end{equation*}
\vspace*{-0.8cm}		
	\begin{equation*}
	{\left(\beta_{17}-\beta_{26}+\beta_{35}-\frac{1}{\sqrt{3}}\beta_{58}  \right)}^2 \leq   | \Vec{a}_2 | | \Vec{a}_6 |,		
	\end{equation*}
\vspace*{-0.8cm}		
	\begin{equation*}
	{\left(\beta_{16}+\beta_{27}+\beta_{34}-\frac{1}{\sqrt{3}}\beta_{48}  \right)}^2 
	\leq   | \Vec{a}_2 | | \Vec{a}_7 |,
	\end{equation*}
\vspace*{-0.8cm}		
	\begin{equation*}
	{\left(\frac{2}{3}\beta_{28}+\frac{2}{\sqrt{3}}\beta_{47}-\frac{2}{\sqrt{3}}\beta_{56} \right)}^2 \leq   | \Vec{a}_2 | | \Vec{a}_8 |,		
	\end{equation*}
\vspace*{-0.8cm}		
	\begin{equation*}
	{\left(\beta_{16}-\beta_{27}-\beta_{34}-\frac{1}{\sqrt{3}}\beta_{48}  \right)}^2  \leq   | \Vec{a}_3 | | \Vec{a}_4 |,		
	\end{equation*}
\vspace*{-0.8cm}		
	\begin{equation*}
	{\left(\beta_{17}+\beta_{26}-\beta_{35}-\frac{1}{\sqrt{3}}\beta_{58}  \right)}^2  \leq   | \Vec{a}_3 | | \Vec{a}_5 |,		
	\end{equation*}
\vspace*{-0.8cm}		
	\begin{equation*}
	{\left(\beta_{14}+\beta_{25}+\beta_{36}-\frac{1}{\sqrt{3}}\beta_{68}  \right)}^2  \leq   | \Vec{a}_3 | | \Vec{a}_6 |,		
	\end{equation*}
\vspace*{-0.8cm}		
	\begin{equation*}
	{\left(\beta_{15}-\beta_{24}+\beta_{37}-\frac{1}{\sqrt{3}} \beta_{78}  \right)}^2  \leq   | \Vec{a}_3 | | \Vec{a}_7 |,		
	\end{equation*}
\vspace*{-0.8cm}		
	\begin{equation*}
	{\left(\beta_{46}-\beta_{57}-\frac{2}{\sqrt{3}}\beta_{18} \right)}^2 \leq   | \Vec{a}_4 | | \Vec{a}_6|,		
	\end{equation*}
\vspace*{-0.8cm}		
	\begin{equation*}
	{\left(\beta_{47}+\beta_{56}+\frac{2}{\sqrt{3}}\beta_{28} \right)}^2 \leq   | \Vec{a}_4 | | \Vec{a}_7 |,		
	\end{equation*}
\vspace*{-0.8cm}		
	\begin{equation*}
	{\left(\beta_{47}+\beta_{56}-\frac{2}{\sqrt{3}}\beta_{28} \right)}^2 \leq   | \Vec{a}_5 | | \Vec{a}_6 |,		
	\end{equation*}
\vspace*{-0.8cm}	
	\begin{equation}
	\label{schwartz2}	
	{\left(\beta_{46}-\beta_{57}+\frac{2}{\sqrt{3}}\beta_{18} \right)}^2 \leq   | \Vec{a}_5 | | \Vec{a}_7 |.		
	\end{equation}

	Moreover, in order to analyze the positivity for the matrix $\bf{A}$, we use for simpliticity  our optimal case   composed by $\beta_{12}$,  $\beta_{28}$, $\beta_{56}$ and $\beta_{47}$, 
	\begin{equation}	
	\bf{D}= \begin{pmatrix}
	-\Gamma&\beta_{12}&0&0&0&0&0&0\\
	\beta_{12}&-\Gamma&0&0&0&0&0&\beta_{28}\\
	0&0&-\Gamma&0&0&0&0&0\\
	0&0&0&-\Gamma&0&0&\beta_{47}&0\\
	0&0&0&0&-\Gamma&\beta_{56}&0&0\\
	0&0&0&0&\beta_{56}&-\Gamma&0&0\\
	0&0&0&\beta_{47}&0&0&-\Gamma&0\\
	0&\beta_{28}&0&0&0&0&0&-\Gamma\\
	\end{pmatrix},
	\end{equation}
then, the matrix $\bf{A}\equiv$$ [a_{kj}]$ is
	\begin{equation}	
	\bf{A}= \begin{pmatrix}
	-\Gamma'&\beta_{12}'&0&0&0&0&0&0\\
	\beta_{12}'&-\Gamma'&0&0&0&0&0&\beta_{28}'\\
	0&0&-\Gamma'&0&0&0&0&0\\
	0&0&0&-\Gamma'&0&0&\beta_{47}'&0\\
	0&0&0&0&-\Gamma'&\beta_{56}'&0&0\\
	0&0&0&0&\beta_{56}'&-\Gamma'&0&0\\
	0&0&0&\beta_{47}'&0&0&-\Gamma'&0\\
	0&\beta_{28}'&0&0&0&0&0&-\Gamma'\\
	\end{pmatrix},
	\end{equation}
	with	
	\begin{equation*}
	\Gamma'=-\frac{2}{3}\Gamma, \hspace{0.4cm}\beta_{12}'=2\beta_{12},
	\end{equation*}
	\vspace*{-1.6cm}
	\begin{equation*}
		\beta_{28}'=\frac{2}{3}\left(\beta_{28}+\sqrt{3}(\beta_{47}-\beta_{56})\right),
	\end{equation*}
	\vspace*{-1.6cm}	
	\begin{equation*}
	\beta_{56}'=\beta_{47}+\beta_{56}-\frac{2}{\sqrt{3}}\beta_{28},
	\end{equation*}
	\vspace*{-1.6cm}
	\begin{equation} 
	\beta_{47}'=\beta_{47}+\beta_{56}+\frac{2}{\sqrt{3}}\beta_{28}.
	\end{equation}
	We get its corresponding eigenvalues: 
	\begin{equation*}
	\lambda_{1,2}=\frac{2}{3}\Gamma \geq 0,
	\end{equation*}
	\vspace*{-1.6cm}
	\begin{equation*}
	\lambda_{3,4}=\frac{1}{3}\left(2\Gamma -3(\beta_{47}+\beta_{56}) \mp 2\sqrt{3}\beta_{28}  \right)\geq 0,
	\end{equation*}
	\vspace*{-1.6cm}
	\begin{equation*}
	\lambda_{5,6}=\frac{1}{3}\left(2\Gamma +3(\beta_{47}+\beta_{56}) \mp 2\sqrt{3}\beta_{28}  \right)\geq 0,
	\end{equation*}
	\vspace*{-1.6cm}
	\begin{equation}
	\begin{split}
	\lambda_{7,8}&=\frac{2}{3}\bigg(\Gamma\mp \big(9\beta_{12}^2+\beta_{28}^2+3(\beta_{47}^2+\beta_{56}^2)\\
	&+2\sqrt{3}\beta_{28}(\beta_{47}-\beta_{56})-6\beta_{47}\beta_{56} \big)^{1/2} \bigg)\geq 0.
    \label{positive_generalcase}
	\end{split}
	\end{equation}
Thus, using the Eqs.~(\ref{positive_generalcase}) and~(\ref{schwartz2}), we can obtain the following individual maximum values for the said $\beta$'s:
$|\beta_{28}|= 1/\sqrt{3}$, $|\beta_{12}|= 1/3$, $|\beta_{56}|= 1/\sqrt{3}$ and $|\beta_{47}|= 1/\sqrt{3}$. 

While, when we set on all 
the aforementioned $\beta$'s together, and following the procedure described in section~\ref{optimalbetas}, we get afterwards: $\beta_{28}=\Gamma/\sqrt{3}$, $\beta_{12}=(\sqrt{2/3})\Gamma/3$ and  $\beta_{56}=-\beta_{47}= \Gamma/3$.


\begin{thebibliography}{0}
\expandafter\ifx\csname natexlab\endcsname\relax\def\natexlab#1{#1}\fi
\expandafter\ifx\csname bibnamefont\endcsname\relax
  \def\bibnamefont#1{#1}\fi
\expandafter\ifx\csname bibfnamefont\endcsname\relax
  \def\bibfnamefont#1{#1}\fi
\expandafter\ifx\csname citenamefont\endcsname\relax
  \def\citenamefont#1{#1}\fi
\expandafter\ifx\csname url\endcsname\relax
  \def\url#1{\texttt{#1}}\fi
\expandafter\ifx\csname urlprefix\endcsname\relax\def\urlprefix{URL }\fi
\providecommand{\bibinfo}[2]{#2}
\providecommand{\eprint}[2][]{\url{#2}}

\end{thebibliography}


\begin{thebibliography}{}
\bibitem{Fukuda01}
S. Fukuda et al. (Super-Kamiokande Collaboration), Phys. Rev. Lett. \textbf{86}, 5651 (2001).
\bibitem{Ahmad02}
Q.R. Ahmad et al. (SNO Collaboration), Phys. Rev. Lett. \textbf{89}, 011302 (2002).
\bibitem{Fukuda98}
Y. Fukuda et al. (Super-Kamiokande Collaboration), Phys. Rev. Lett. \textbf{81}, 1562 (1998).
\bibitem{Kajita16}
T. Kajita et al. (Super-Kamiokande Collaboration), Nucl. Phys. B \textbf{908}, 14 (2016).
\bibitem{Araki05}
T. Araki et al. (KamLAND Collaboration), Phys. Rev. Lett. \textbf{94}, 081801 (2005).
\bibitem{An12}
F.P. An et al. (Daya Bay Collaboration), Phys. Rev. Lett. \textbf{108}, 171803 (2012).
\bibitem{Adamson14}
P. Adamson et al. (MINOS Collaboration), Phys. Rev. D \textbf{77}, 072002 (2008).
\bibitem{Ahn12}
J.K. Ahn et al. (RENO Collaboration), Phys. Rev. Lett. \textbf{108}, 191802 (2012).
\bibitem{Abe12}
Y. Abe et al. (Double CHOOZ Collaboration), Phys. Rev. Lett. \textbf{108}, 131801 (2012).
\bibitem{McDonald05}
A.B. McDonald et al. (SNO Collaboration), Phys. Scripta T \textbf{121}, 29 (2005).

\bibitem{Gasperini88}
M.~Gasperini,
Phys.\ Rev.\ D {\bf 38},  2635 (1988).
\bibitem{Halprin91} 
A.~Halprin and C.~N.~Leung,
Phys.\ Rev.\ Lett.\  {\bf 67}, 1833 (1991).

\bibitem{Gago:1999hi} 
  A.~M.~Gago, H.~Nunokawa and R.~Zukanovich Funchal,
  Phys.\ Rev.\ Lett.\  {\bf 84}, 4035 (2000)

\bibitem{Colladay:1996iz} 
  D.~Colladay and V.~A.~Kostelecky,
  Phys.\ Rev.\ D {\bf 55}, 6760 (1997),
\bibitem{Coleman:1998ti} 
  S.~R.~Coleman and S.~L.~Glashow,
  Phys.\ Rev.\ D {\bf 59}, 116008 (1999).
\bibitem{Coleman:1997xq} 
  S.~R.~Coleman and S.~L.~Glashow,
  Phys.\ Lett.\ B {\bf 405}, 249 (1997)
\bibitem{Colladay:1998fq}
  D.~Colladay and V.~A.~Kostelecky,
  Phys.\ Rev.\ D {\bf 58}, 116002(1998).
 

\bibitem{Adamson:2008aa} 
  P.~Adamson {\it et al.} [MINOS Collaboration],
  Phys.\ Rev.\ Lett.\  {\bf 101}, 151601 (2008)
\bibitem{Adamson:2010rn} 
  P.~Adamson {\it et al.} [MINOS Collaboration],
  Phys.\ Rev.\ Lett.\  {\bf 105}, 151601 (2010)
\bibitem{AguilarArevalo:2011yi} 
  A.~A.~Aguilar-Arevalo {\it et al.} [MiniBooNE Collaboration],
 Phys.\ Lett.\ B {\bf 718}, 1303 (2013).
\bibitem{Li:2014rya} 
  Y.~F.~Li and Z.~h.~Zhao,
  Phys.\ Rev.\ D {\bf 90}, no. 11, 113014 (2014).

\bibitem{Jacobson:2003wc} 
  M.~Jacobson and T.~Ohlsson,
  Phys.\ Rev.\ D {\bf 69}, 013003 (2004)
\bibitem{Ellis}
J. Ellis, N. E. Mavromatos and D. V. Nanopoulos, Phys. Lett. \textbf{B293}, 37 (1992); Int. J. Mod. Phys. \textbf{A11}, 1489 (1996).
\bibitem{Benattistrings} 
F. Benatti and R. Floreanini, Ann. of Phys. \textbf{273}, 58 (1999)
\bibitem{Hawking1}
S. Hawking, Comm. Math. Phys. \textbf{87}, 395 (1983); Phys. Rev. D \textbf{37}, 904 (1988); Phys. Rev. D \textbf{53}, 3099 (1996); S. Hawking and C. Hunter, Phys. Rev. D \textbf{59}, 044025 (1999).






\bibitem{Benatti00}
F. Benatti and R. Floreanini, JHEP0002, 32 (2000).
\bibitem{Benatti01}
F. Benatti and R. Floreanini, Phys. Rev. D \textbf{64}, 085015 (2001).
\bibitem{Lisi00}
E. Lisi, A. Marrone and D. Montanino, Phys. Rev. Lett. \textbf{85}, 1166 (2000).
\bibitem{Gago01a}
A.M. Gago, E.M. Santos, W.J.C. Teves and R. Zukanovich Funchal, Phys. Rev. D \textbf{63}, 073001 (2001).
\bibitem{Gago02a}
A.M. Gago, E.M. Santos, W.J.C. Teves and R. Zukanovich Funchal, arXiv:0208166.
\bibitem{Barenboim:2004wu} 
 G.~Barenboim and N.~E.~Mavromatos,
 JHEP {\bf 0501}, 034 (2005).
\bibitem{Fogli07}
G. L. Fogli, E. Lisi, A. Marrone, D. Montanino and A. Palazzo, Phys. Rev. D \textbf{76}, 033006 (2007).
\bibitem{Farzan08}
Y. Farzan, T. Schwetz and A.Y. Smirnov, JHEP0807, 067 (2008).
\bibitem{Oliveira13}
R.L.N. Oliveira and M.M. Guzzo, Eur. Phys. J. C \textbf{73}, 2434 (2013).
\bibitem{Carpio:2017nui} 
  J.~A.~Carpio, E.~Massoni and A.~M.~Gago,
  Phys.\ Rev.\ D {\bf 97}, no. 11, 115017 (2018).
\bibitem{Wald80}
R. M. Wald, Phys. Rev. D \textbf{21}, 2742 (1980). 
\bibitem{Mavromatos:2009ww} 
  N.~E.~Mavromatos,
  J.\ Phys.\ Conf.\ Ser.\  {\bf 171}, 012007 (2009).

\bibitem{Liu97} 
Y.~Liu, L.~z.~Hu and M.~L.~Ge,
Phys.\ Rev.\ D {\bf 56}, 6648 (1997).			

\bibitem{EllisMavro96} 
J.~R.~Ellis, J.~L.~Lopez, N.~E.~Mavromatos and D.~V.~Nanopoulos,
Phys.\ Rev.\ D {\bf 53}, 3846 (1996); J.~R.~Ellis, N.~E.~Mavromatos and D.~V.~Nanopoulos,
Phys.\ Rev.\ D {\bf 63}, 024024 (2001). 





\bibitem{cdr}
R.~Acciarri {\it et al.} [DUNE Collaboration], arXiv:1512.06148 (2015).
\bibitem{ancillaryfiles}
T.~Alion {\it et al.} [DUNE Collaboration], arXiv:1606.09550 (2016).
\bibitem{Pontecorvo:1957cp} 
  B.~Pontecorvo,
  Sov.\ Phys.\ JETP {\bf 6}, 429 (1957)
  [Zh.\ Eksp.\ Teor.\ Fiz.\  {\bf 33}, 549 (1957)].
\bibitem{Maki:1962mu} 
  Z.~Maki, M.~Nakagawa and S.~Sakata,
  Prog.\ Theor.\ Phys.\  {\bf 28}, 870 (1962).
\bibitem{Nufit}
http://www.nu-fit.org



\bibitem{marvin} 
  M.~V.~Ascencio-Sosa, A.~M.~Calatayud-Cadenillas, A.~M.~Gago and J.~Jones-Pérez,
  Eur.\ Phys.\ J.\ C {\bf 78}, no. 10, 809 
  arXiv:1805.03279 (2018).

\bibitem{MasudDUNE} 
  M.~Masud, M.~Bishai and P.~Mehta,
  arXiv:1704.08650 (2017).

\bibitem{Berryman} 
  J.~M.~Berryman, A.~de Gouvêa, K.~J.~Kelly and A.~Kobach,
  Phys.\ Rev.\ D {\bf 92}, no. 7, 073012 (2015)



\bibitem{globes1}
P.~Huber, M.~Lindner and W.~Winter,  Comput.\ Phys.\ Commun.\  {\bf 167}, 195 (2005).
\bibitem{globes2}
P.~Huber, J.~Kopp, M.~Lindner, M.~Rolinec and W.~Winter,  Comput.\ Phys.\ Commun.\  {\bf 177}, 432 (2007).
\bibitem{nusquids}
C.~A.~Arg\"{u}elles Delgado, J.~Salvado and C.~N.~Weaver, Comput.\ Phys.\ Commun.\  {\bf 196}, 569 (2015), arXiv:1412.3832.
\bibitem{Hernandez16}
S. Hern\'andez, Magister t\'esis, Pontificia Universidad Cat\'olica del Per\'u, 2016.
	\end{thebibliography}
\end{document}